\documentstyle{article}
\begin{document}
\title{\Large\bf  Ground State of the Double Exchange Model \\}
\author{\large
        Liang-Jian Zou$^{a}$, Qing-Qi Zheng$^{a,b}$, H. Q. Lin$^{c}$ \\
{\it$^a$Institute of Solid State Physics, Academia Sinica,
        P.O.Box 1129, Hefei 230031, China} \\
{\it$^b$State Key Lab of Magnetism,
    Institute of Physics, Academia Sinica, Beijing, China} \\
{\it$^c$Department of Physics, Chinese University of Hong Kong,
        Shatin, N.T. Hong Kong, China } \\ }
\date{\today}
\maketitle
\large
\begin{center}
{\bf Abstract}
\end{center}
We investigate the electronic correlation effect on the ground-state
properties of the double exchange model for manganites by using a
semiclassical approach and the slave-boson technique.
It is shown that magnetic field has a similar effect on the canted angle
between manganese spins as doping concentration does, and
the canted angle exhibits weak dependence on the Coulomb interaction.
The possibility of phase separation in the present model is also discussed.
In the slave-boson saddle-point approximation in the ferromagnetic metallic
regime, the dependence of the magnetization and the Curie temperature
on the doping concentration exhibits maxima near 1/3 doping.
These results agree with experimental data and suggest that the electronic
correlation plays an important role for understanding the ground-state
properties of manganites.  \\

%\vspace{1cm}

\vspace{1cm}

\newpage

  It is essential to clarify the ground state and magnetic phase diagram for
elucidating the microscopic mechanism of the colossal magnetoresistance (CMR)
in lanthanum manganese. The ground state and the magnetic phase diagram of
lanthanum manganese at low doping concentration in low temperature
are still controversial, though some efforts [1-8] have been devoted to it.
In 1950, Zener [5] proposed a double exchange (DE) model to explain
the electrical conduction and the ferromagnetism (FM) of doped lanthanum
manganese. Later Anderson and Hasegawa [6] derived the DE energy for a pair
of Mn ions and showed that in such a system,
the DE interaction tends to align the spins of Mn ions parallel and the DE
energy is proportional to cos($\theta_{ij}/$2), not to cos($\theta_{ij}$) as
in the Heisenberg model, here $\theta_{ij}$ denotes the angle between spins
{\bf S}$_{i}$ and {\bf S}$_{j}$. In 1960, De Gennes [7] generalized their
results to the case with finite doping concentration.
He assumed that the total DE energy is proportional to $cos(\theta_{ij}/2)$,
and showed that the Mn spins in the case of finite doping are
ferromagnetically ordered but canted by an angle $\theta$ that depends on
the carrier concentration before a critical concentration x$_{c}$.
Since then the concept of canted ferromagnet or antiferromagnet was accepted,
but not confirmed definitively by early experiments [8].
In recent experiments some researchers declared that there exists canted
structure [4,9], but negative results were also reported.
Schiffer et al. [1] reported that at low doping (0$<x<$0.2),
La$_{1-x}$Ca$_{x}$MnO$_{3}$ is ferromagnetically ordered, whereas
Jonker and Van Santan's early report [2] suggested an antiferromagnetic order.
Martin et al. [4] showed that La$_{1-x}$Sr$_{x}$MnO$_{3}$ is spin-canted
for 0$<x<$0.1, and ferromagnetic ordered for 0.1$<x<$0.2.
These reports on the low-temperature low-doping magnetic phase diagram do not
agree with each other. Thus it is necessary to study the DE model in details
to clarify the magnetic structure in low-doping regime.

    Both the early and the recent experiments [1-4] have shown that in
La$_{1-x}$R$_{x}$MnO$_{3}$ (R=Ca, Sr), the magnetization and
the Curie temperature exhibit maxima around x=1/3.
Theoretically, these observations have not been explained satisfactory.
Varma [10] estimated that the maximum of the Curie
temperature appears at 1/2 doping, Xing and Shen [11] also showed that the
zero-temperature magnetization reaches its maximum near 1/2 doping.
Another interesting problem is how the magnetic field affects the magnetic
structure, since the resistivity of doped lanthanum manganeses is changed
by several orders of magnitude under the external magnetic field,
such a huge change might be related to the variation of the
magnetic structure modulated by magnetic field.
Furthermore, the role of electronic correlation was taken into account
lightly in previous studies [5-8], since in the primary DE model it only
includes the Hund's coupling between conduction electrons and the core spins
but not the Coulomb interaction among conduction electrons.
A clear picture of the ground state magnetic properties is needed in
order to have a coherent understandings of these phenomena in manganites.
In the present paper, we first derive the DE energy in the presence of the
Coulomb interaction and the magnetic field, then discuss doping dependence of
the mean-field ground state energy, the magnetization and the Curie
temperature in ferromagnetic metallic regime in the strong correlation limit.
\\

\noindent {\bf I. Diagonalization in Momentum Space}.

The electronic states in doped lanthanum manganese have been depicted in many
papers [5-8,12], in the presence of Coulomb interaction and magnetic field,
the model Hamiltonian can be written as a summation of two parts: the double
exchange interaction $H_{DE}$ and the superexchange interaction $H_{m}$
\[    H=H_{DE}+H_{m}   \]
\begin{equation}
  H_{DE} =\sum_{<ij>\sigma} t_{ij} d^{\dag}_{i \sigma}d_{j \sigma}
 +\frac{U}{2} \sum_{i\sigma} n_{i\sigma}n_{i\bar{\sigma}}
 - J_{H} \sum_{i\mu \nu} {\bf S}_{i} \cdot d^{\dag}_{i \mu}
        \mbox{\boldmath $\sigma$}_{\mu \nu} d_{i\nu}
\end{equation}
\begin{equation}
   H_{m} = - g\mu_{B}B\sum_{i} S^{z}_{i}
       +\sum_{<ij>}{\it A}_{ij} {\bf S}_{i} \cdot {\bf S}_{j} ~,
\end{equation}
where the three d electrons of Mn ions are in the t$_{2g}$ state at site
R$_{i}$ and they form a localized core spin ${\bf S}_{i}$,
d$^{\dag}_{i \sigma}$ creates a mobile electron in the e$_{g}$ band at site
R$_{i}$ with spin $\sigma$, t$_{ij}$ denotes effective hopping matrix element
of the mobile electrons to its nearest neighbor, U denotes the on-site Coulomb
interaction among mobile electrons, and $J_{H}$ represents the Hund's
coupling between the local spins and the mobile electrons, $J_{H} \gg zt/S$
as required by the DE mechanism.
In Eqs.(1) and (2), $< \cdots >$ indicates that only the nearest neighbor
interaction is considered. In Eq.(2), g$\mu_{B}$ represents the effective
magnetic moment of local spin, {\bf B} represents the external magnetic field,
and the last term represents the superexchange interaction between Mn ions,
{\it A}$_{ij}$ denotes the superexchange interaction constant which is
negative (-$A^{'}$) for ${\bf R}_{i}$ and ${\bf R}_{j}$ on the {\it ac} plane
and positive ($A$) for ${\bf R}_{i}$ and ${\bf R}_{j}$ on the {\it b}-axis.
The Jahn-Teller effect and electron-phonon interaction are not included here.

To start, we assume ${\bf S}_{i}$ are classical spins in this section, which
corresponds to the following substitution:
\begin{equation}
S^{z}_{i} = S cos(\theta),~~~ S^{\pm}_{i}=S e^{\pm i{\bf Q \cdot R}_{i}}
sin(\theta)
\end{equation}
here {\bf Q}=(0,$\pi/b$,0), $\theta$ is the canted angle and 2$\theta$ the
angle between two spins. For small doping $LaMnO_{3}$, the carrier is hole,
translating the electron representation into hole representation, \( h^{+} \)
the model Hamiltonian can then be expressed in momentum space:
\begin{eqnarray}
  H &=& \sum_{k}[-g\mu_{B}B S cos(\theta)
       - 4{\it A^{'}}S^{2} + 2{\it A}S^{2}cos(2\theta)] \nonumber \\
    &+& \sum_{k\sigma} [ ( -\epsilon_{k} + U<n_{\bar{\sigma}}>
                          +\sigma J_{H}S )h^{\dag}_{k\sigma}h_{k\sigma}
       + J_{H}S sin(\theta) ( h^{\dag}_{k+Q\uparrow} h_{k\downarrow}
                            + h^{\dag}_{k\downarrow} h_{k+Q\uparrow} ) ]
\label{H_k}
\end{eqnarray}
where $\epsilon_{k\sigma}=zt\gamma_{k}$ denotes dispersion of holes,
$\gamma_{k} = (1/3) ( cos k_x + cos k_y + cos k_z ) $
is the structure factor. In this section the Coulomb
interaction is treated by the Hatree-Fock approximation.
Diagonalization of the hole part in Eq. (4) gives rise to two subbands:
\begin{equation}
   E_{k\sigma} = U<n_{\bar{\sigma}}> \pm
\sqrt{\epsilon_{k\sigma}^{2}+(J_H S)^{2}+2J_H S\epsilon_{k\sigma}cos(\theta)}
\label{E_k}
\end{equation}
A similar expression has been obtained by Dimashko et al. [13] to address
the phase separation issue for high-temperature superconductivity in the
limit of 2zt/J$_{H}$S $>>$1 with U=0.
Inoue et al. [14] also obtained similar expression in the DE limit
and suggested that a spiral state may be more stable than the canted state
in La$_{1-x}$R$_{x}$MnO$_{3}$, but they did not consider the effect of the
Coulomb interaction on the ground state and the possibility of
phase separation. Later we will show that the Coulomb correlation can not be
neglected.

To explore the ground-state properties of lanthanum manganeses, we are only
interested of the lower subband of (5).
In the DE model, zt/J$_{H}$S is a small quantity and we can expand E$_{k}$
to the linear term of zt/J$_{H}$S. At zero temperature,
the ground-state energy of the system with uniform doping concentration x is:
\begin{equation}
  E_{G} = NS[-g\mu_{B}Bcos(\theta)-4A^{'}S+2AScos(2\theta)]
        + \sum_{k\sigma}^{ k_{F}}
        [ U<n_{\bar{\sigma}}>-J_{H}S-\epsilon_{k\sigma}cos(\theta) ] ~,
\end{equation}
where $N$ is the total number of core spins.
The summation of the mean occupation over spin is the carrier concentration,
i.e., $\sum_{k \sigma}^{k_F} <n_{k\sigma}>$=x, the Fermi wave vector is
$k_{F}$.

Minimizing the total energy with respect to $\theta$ gives rise to
the canting angle,
\begin{equation}
        cos(\theta)=\frac{g\mu_{B}BS + 2zt \alpha}{8AS^{2}}~,~~~~
                    \alpha=\frac{2}{N} \sum_{k}^{k_{F}} \gamma(k) ~.
\end{equation}
For small doping concentration, $x \ll 1$, $\alpha$ depends on doping
concentration, we have
\[ k_F^3 = 3 \pi^2 x ~,\]
\[ \alpha = x [ 1- (3\pi^{2}x)^{2/3}/10]   ~. \]
in a three-dimensional isotropic lattice system.
This result is slightly different from that of [7], due to the lattice effect.
In the absence of the external magnetic field ($B=0$) for very small doping,
$\alpha \approx x$, the critical hole density
for the system evolving from canted antiferromagnet into ferromagnet is
x$_{c}=4 A S^2 / zt$, this result is similar to that of Ref [7].
Both the present result (in the limit zt/$J_{H}S$ $<<1$) and that of [13]
(in the limit zt/J$_{H}S$ $>>1$) have shown that the ground state is
antiferromagnetic in the zero doping limit when there is no external magnetic
field, so it is reasonable to expect that the ground state is always
antiferromagnetic for all values of zt/J$_{H}S$ in pure lanthanum manganites.

Furthermore the present theory contains some more interesting results.
First, the influence of the external magnetic field
on the magnetic structure can be discussed for almost pure
lanthanum manganites (x $approx$ 0), the effect of magnetic field is similar
to that of
doping, the cosine of canted angle linearly increases with magnetic field.
At a certain critical value B$_{c}$ =8AS/g$\mu_{B}$, the external magnetic
field exceeds the superexchange field, all spins tend to align paralleled, 
the ferromagnetic alignment of local spins are in favor of the motion of
holes, so the system may exhibit large decrease in resistivity, however the
critical field may be as high as hundreds of T, so it would not like that
%for $4AS^2 = 4.84 mev$, $B_c=(2S/g)(4AS^2/\mu_B)=(2S/g)(4.84/5.7884)10^2 T$.
the metal-insulator transition induced by the external magnetic field
causes the CMR effect.
Second, in the Hatree-Fock approximation and expanding E$_{k\sigma}$ (see (5))
to the second order of (2zt/J$_{H}$S), one can find that the canted angle
weakly depends on the Coulomb interaction U,
so the consideration of the Coulomb correlation in the mean-field
approximation does not change canted angle significantly.
This is only attribute to the fact that treatment of the electronic correlation
in the Hatree-Fock approximation is rather rough.

One conclusion of the above discussion is that manganites with uniform hole
concentration is spin canted at low doping. However, Schiffer {\it et al.}'s
report [1] on a low-doping phase diagram suggests ferromagnetic ordering.
This may have two possible reasons:
the first is that the oxygen content in La$_{1-x}$Ca$_{x}$MnO$_{3+\delta}$ is
not exactly stoichiometric ($\delta \neq 0)$, so the ferromagnetic component
arising from the DE interaction plays a role;
the second is that phase separation might take place, holes aggregate into
a ferromagnetic droplet, so the ferromagnetic ground-state emerges.
In the following, we briefly discuss the possibility of phase separation
in the DE model (2zt/J$_{H}$S $<<$1), as contrast to the usual
{\it s-f} model (2zt/J$_{H}$S $>>$1).
After the holes aggregate into droplets from the antiferromagnetic background,
in the absence of external magnetic field,
the energy densities e(x) in the hole-rich phase at hole density x is e(x):
\begin{equation}
  e(x) = S[-4A^{'}S+2AScos(2\theta)]
       + [\frac{Ux^{2}}{2}-xJ_{H}S- 2zt\alpha cos(\theta)] ~,
\end{equation}
and leaving the hole-free antiferromagnetic background with energy density
e(0), $e(0) =-2S^{2}[2A^{'}+A]$, here magnetic field B=0.
Let \( n_{h} \) be the total number of hole, then the number of sites
occupied by the hole-rich phase is n$_{h}$/x. N is the number of sites of the
whole system. Then the total energy of the two-phase state is:
\begin{equation}
  E(x)=-2NS^{2}(2A^{'}+A)-2n_{h}J_{H}S+n_{h}[\frac{4AS^{2}cos^{2}(\theta)}{x}
             +\frac{Ux}{2}- \frac{4zt\alpha cos(\theta)}{x} ]
\end{equation}
For very low hole concentration, $\alpha \approx x$, one has:
\begin{equation}
  E(x) =  \left\{ \begin{array}{ll}
 {\it const}+n_{h}(\frac{U}{2}-\frac{(zt)^{2}}{AS^{2}}) x  &  x < x_{c} \\
 {\it const}+n_{h}(\frac{4AS^{2}}{x}+\frac{Ux}{2}) & x \geq x_{c}
    \end{array} \right.
\end{equation}
One finds that the presence of a strong on-site Coulomb interaction may
prevent phase separation, however, if U is smaller than a critical
value U$_{c}$,
\begin{equation}
   U_{c} = \frac{(2zt)^{2}}{AS^{2}}
\end{equation}
the two-phase energy has a minimum at density
\[    x=(\frac{8AS^{2}}{U})^{1/2}  ~, \]
so the phase separation into ferromagnetic droplet takes place at sufficiently 
low density $x <x_{0}$. When $U>U_c$, the E(x) dependence is monotonous
[E'(x)>0], so there is no phase separation at all.
For typical parameters in \( La_{1-x}Ca_{x}MnO_{3}, 4AS^{2}=4.84 meV  [15] \),
and by the electronic structure calculation of the local density functional
technique, we find that 2zt=0.5 eV, therefore $x_0 \approx 0.007 $, which is
much smaller than the critical concentration $x_c (\approx 0.1)$.
This may address the experimental observation in Ref. 1. Further experiment
is expected.\\

\noindent {\bf II. A Mean-Field Solution.}

In the $La_{1-x}R_{x} MnO_{3}$ system, there exists Mn$^{+3}$ and Mn$^{+4}$
ions. Due to strong Hund's coupling and Coulomb interactions [16],
the Mn$^{+2}$ ions are excluded, i.e., double occupancy in
the $e_{2g}$ orbital is prohibited. The hopping integral $t$ is far less than
the on-site Coulomb interaction and the Hund's coupling, so it is reasonable to
take U as infinity to exclude the appearance of Mn$^{+2}$ in manganites, or
the double occupation. In the limit of large Coulomb interaction,
the constraint of no double occupancy at site $R_{i} $ can be enforced by
introducing auxiliary fermions [17], f$_{i\sigma}$, and bosons, b$_{\sigma}$,
where $f^{\dag}_{i \sigma}$ creates a slave fermion with spin $\sigma$
when site $R_{i}$ is occupied, while $b^{\dag}_{i}$ creates a boson (hole)
at R$_{i}$ when it is unoccupied.
Thus $d_{i\sigma}=f_{i\sigma}b^{\dag}_{i}$ and the model Hamiltonian can be
rewritten as:
\begin{equation}
   H =\sum_{<ij>\sigma} t_{ij} f^{\dag}_{i \sigma}f_{j\sigma}b_{i}b^{\dag}_{j}
   -J_{H} \sum_{i\mu\nu} {\bf S}_{i} \cdot f^{\dag}_{i \mu}
      {\bf \sigma}_{\mu \nu} f_{i\nu}
       +\sum_{<ij>}{\it A}_{ij} {\bf S}_{i} \cdot {\bf S}_{j}
      +\sum_{i} \epsilon_{d} (\sum_{\sigma}f^{\dag}_{i \sigma}f_{i \sigma}+
      b^{\dag}_{i}b_{i}-1) ~
\end{equation}
where $\epsilon_{d}$ is the energy shift of the d-electron with respect to
the original energy level, the other parameters are the same as in Eq. (1).

   In the static (or saddle-point) approximation, the boson field is replaced
by its mean value and assumed
to be independent of R$_{i}$, $<b^{\dag}_{i}>=<b_{i}>=b^{1/2}$, and one can
obtain the mean-field equations by taking derivatives with respect
to $\epsilon_{d}$ and $b$:
\begin{equation}
   \sum_{\sigma} <f^{\dag}_{i \sigma}f_{i \sigma}> =1- b ~
\end{equation}
\begin{equation}
\epsilon_{d} = - 2 t \sum_{\delta} <f^{\dag}_{i \sigma}f_{i+\delta \sigma}> ~
\end{equation}
Physically, b gives rise to the mean carrier (hole) concentration on every
site (see Eq.(13)). If the localization effect of the carriers is neglected,
$b$ corresponds to the doping concentration, $x$.
Since spin components are relevant to the carrier concentration and the
spin-dependent energy should be included in the mean value of
fermion propagator, the spin configuration and the carrier
concentration must be determined self-consistently.

  The mean values in Eqs.(13) and (14) can be obtained from the fermion
propagators, $G_{\sigma}$(ij;$\omega)$:
\begin{equation}
   G_{\sigma}(ij;\omega) =\sum_{k}
   1/[w-\epsilon_{d}-2ztb\gamma_{k}+\sigma J_{H} S^{z}_{Q}]
        e^{{\bf k} \cdot ({\bf R}_{i}-{\bf R}_{j})}
\end{equation}
where $S^{z}_{Q}$ denotes the $z$-component of the spin with wave vector
${\bf Q}$: ${\bf Q}=0$ corresponds to ferromagnetic order, ${\bf \pi}$ to
antiferromagnetic order, and values between 0 and $\pi$ to canted structures.
Then the self-consistent equations at zero temperature are:
\begin{equation}
   1-b=-\frac{1}{\pi N}
\sum_{k\sigma} \int^{\epsilon_{F}} d\omega {\it Im}
\frac{1}{\omega+i\eta-\epsilon_{d}-2ztb\gamma_{k}+\sigma J_{H} S^{z}_{Q}} ~
\end{equation}
and:
\begin{equation}
  \epsilon_{d}= \frac{4zt}{\pi N} \sum_{k\sigma} \gamma(k)
     \int^{\epsilon_{F}} d\omega {\it Im}
     \frac{1}{\omega+i\eta-\epsilon_{d}+2ztb\gamma_{k}+\sigma J_{H} S^{z}_{Q}},
\end{equation}
where $\epsilon_{F}$ is the Fermi energy.
Accordingly, we can obtain the mean value of $\ < S^{z}_{Q} \ >$, the energy
shift $\epsilon_{d}$ and the ground state energy E$_{g}$ for doping
concentration, $b (=x)$, at zero temperature.

  In the present section we are interested in the ferromagnetic metallic regime
of La$_{1-x}$Ca$_{x}$ MnO$_{3}$ (0.2$<x<0.5$) system, where the $z$-component of
the spin, ${\it S}^{z}$, is the same at all the sites and is independent of
the wave vector ${\bf Q}$. In the ferromagnetic metallic regime, the carrier is
completely spin-polarized due to the strong Hund's coupling,
the density of states of the fermion may take a simple form:
\begin{equation}
  {\rho} (\epsilon) = \left\{ \begin{array}{ll}
1/2bD & |\epsilon-\epsilon_{d}+ J_{H}<S^{z}>| < bD
\\ 0 & |\epsilon-\epsilon_{d}+J_{H}<S^{z}>| >bD  \end{array}
\right.
\end{equation}
where $2bD$ is the bandwidth of fermion, the solutions of the self-consistent
mean-field equations give rise to the energy shift, $\epsilon_{d}$,
\begin{equation}
   \epsilon_{d} = Db(1-b)=D(1-n^{f})n^{f}
\end{equation}
and the local spin moment:
\begin{equation}
   <S^{z}> =(-\epsilon_{F}+2bD-3b^{2}D)/J_{H}
\end{equation}
at zero temperature. An interesting result is that there is an optimized
doping for the local spin moment, or the magnetization.
From Eq.(20), one finds that the local spin will have a maximum at $b=1/3$.
Since the magnetization $M$ is proportional to $<S^{z}>$, and as
mentioned above, $b$ corresponds to the hole or doping
concentration, so one could expect that the magnetization exhibits a maximum
around the doping concentration of 1/3, which agrees with experimental
observations in La$_{1-x}$Ca$_{x}$MnO system [2,3]. Furthermore, one could show
by a simple analysis that the Curie temperature also reaches to its maximum
around 1/3 doping, which is in agreement with experiments [1,4], and different
from the theoretical results in Refs. 10,11.

   In the preceding discussion, the electron localization character resulting
from the disorder effect in doping is not taken into account, and if it is
taken
into account, we could expect that optimizing doping concentration for
magnetization and Curie temperature may not be at x=1/3 precisely.
It could be a little larger than 1/3. Therefore the complete consideration of
the electron correlation is important to understand the ground state properties
of CMR materials.

     To summarize, external magnetic field has similar effect on the canted
angle of the manganese spins as the doping concentration,
the phase separation may take place in doped manganites. The mean-field
magnetization and the Curie temperature reach maxima near 1/3 doping. \\

\noindent {\bf Acknowledgement}
   L.-J. Zou thanks the invitation of the International Centre of Theoretical
Physics (ICTP) in Trieste, Italy.
This work is partly supported by the Grant of NNSF of China and the Grant
of Chinese Academy of Science,
and by the Direct Grant for Research from the Research Grants Council (RGC)
of the Hong Kong Government.

\newpage
\begin{center}
REFERENCES
\end{center}
\begin{enumerate}
\item P. Schiffer, A. P. Ramirez, W. Bao and S-W. Cheong,
       {\it Phys. Rev. Lett.} {\bf 75}, 3336 (1995).
\item G. H. Jonker and J. H. Van Sauten, {\it Physica}, {\bf 16}, 337 (1950);
      J. H. Van Sauten and G. H. Jonker, {\it ibid}, {\bf 16}, 599 (1950).
\item E. Q. Wollen and W. C. Koeller, {\it Phys. Rev.}, {\bf 100}, 545 (1955).
\item M. C. Martin, G. Shirane, Y. Erdoh, K. Hirota, Y. Moritomo and Y. Tokura,
       {\it Phys. Rev.} {\bf B53}, 14285 (1996).
\item C. Zener, {\it Phys. Rev.}, {\bf 81}, 440 (1951); {\bf 82}, 403 (1951).
\item P. W. Anderson and H. Hasegawa, {\it Phys. Rev.}, {\bf 100}, 675 (1955).
\item P. G. De Gennes, {\it Phys. Rev.}, {\bf 100}, 564 (1955); {\bf 118}, 141
      (1960).
\item K. Kubo and N. Ohata,  {\it J. Phys. Soc. Jpn.} {\bf 33}, 21 (1972).
\item H. Yoshizawa, H. Kawano, Y. Tomioka and Y. Tokura, {\it Phys. Rev.}
      {\bf B52}, 13145 (1995).
      H. Kawano, R. Kajimoto, M. Kubota and H. Yoshizawa, {\it Phys. Rev.}
      {\bf B53}, 2202 (1996).
\item C. M. Varma, {\it Phys. Rev.} {\bf B54} 7328 (1996).
\item D. Y. Xing and Sheng Li, {\it unpublished}
\item J. M. Coey, M. Viret, J. Ranno and K. Ounadjela, {\it Phys. Rev. Lett.},
      {\bf 75}, 3910 (1995).
\item Y. A. Dimashko and A. L. Alistratov, {\it Phys. Rev.}, {\bf 50}(2),1162
      (1996)
\item J. Inoue and S. Maekawa, {\it Phys. Rev. Lett} {\bf 74}, 3407 (1995).
\item K. Hirota, N. Kaneko, A.Nishizawa and Y. Endoh, {\it J. Soc. Phys. Jap.}
      {\bf 65}, 3736 (1996)
\item S. Satpathy, Z. S. Povovic and F. R. Vukajlovic,
      {\it J. Appl. Phys.} {\bf 79} 4555 (1996).
\item P. Coleman  {\it Phys. Rev.}, {\bf B29}, 3055 (1984)
\end{enumerate}

\end{document}